\documentclass[aps,pre,twocolumn,longbibliography,showpacs,amsfonts,amssymb,amsmath]{revtex4-1}
\usepackage{graphicx,xcolor,soul}
\newcommand{\mean}[1]{\langle #1\rangle}
\renewcommand{\vec}[1]{\mathbf{#1}}

\newcommand{\x}{\vec r}
\newcommand{\X}{\vec R}
\newcommand{\gd}{\dot{\gamma}}
\newcommand{\re}{R_\text{ee}}
\newcommand{\zc}{z_\text{c}}
\newcommand{\al}{\alpha}

\newcommand{\sig}{\sigma}

\newcommand{\ea}{\textit{et al. }}
\newcommand{\mat}[1]{\boldsymbol{#1}}

\begin{document}

\title{Non-Equilibrium Markov State Modeling of the Globule-Stretch Transition}

\author{Fabian Knoch and Thomas Speck}
\affiliation{Institut f\"ur Physik, Johannes Gutenberg-Universit\"at Mainz,
  Staudingerweg 7-9, 55128 Mainz, Germany}

\begin{abstract}
  We describe a systematic approach to construct coarse-grained Markov state models from molecular dynamics data of systems driven into a non-equilibrium steady state. We apply this method to study the globule-stretch transition of a single tethered model polymer in shear flow. The folding/unfolding rates of the coarse-grained model agree with the original detailed model. We demonstrate that the folding/unfolding proceeds through the same narrow region of configuration space but along different cycles.
\end{abstract}


\maketitle

\section{Introduction}

Computer simulations have developed into a powerful tool to predict and optimize material properties. However, even given the ever increasing computational power, relevant time and length scales, in particular for biological and synthetic macromolecules~\cite{karp02,kame11}, will remain prohibitively long for fully atomistic simulations, and multiscale methods are crucial to make progress. While structure-based coarse-graining~\cite{mull02} has been quite successful~\cite{harm09,sale16}, challenges remain like transferability (\emph{e.g.}, from bulk to surfaces), and in particular the correct treatment (or \emph{a posteriori} deduction) of the materials \emph{dynamical} properties through the coarse-grained dynamics. For the latter, Markov state models (MSMs)~\cite{noe2008,noe2009,bowm11,prinz2011} have been used successfully to tackle the evolution of large proteins towards their native state, bridging the gap from molecular dynamics on nanoseconds to folding on milliseconds~\cite{dobs03,noe2015}. The dynamics of MSMs is a discrete-time master equation with transition probabilities obey detailed balance on a network of long-lived (metastable) mesostates.

Here we describe a systematic approach to construct MSMs with dynamics that break detailed balance. We employ this method to study a model polymer in shear flow. The rheology of dilute flexible polymers has been studied extensively due to their fundamental and practical relevance~\cite{lars05}. Examples include biomolecules such as the von Willebrand factor in blood plasma and DNA in steady shear flow~\cite{smit99}. The shear drag can overcome the entropic forces favoring coiled or globular configurations and stretch the polymers, which might be a continuous or even discontinuous transition~\cite{genn74}. Motion of DNA tethered to a planar surface has been described as \emph{cyclic} in experiments~\cite{doyl00} and computer simulations~\cite{delgado2006,zhan09}.

Driving a system away from thermal equilibrium implies a non-vanishing entropy production. In a steady state, this entropy is produced exclusively in \emph{cycles} since there are no sources and sinks for the probability. This poses novel challenges for constructing coarse-grained models in non-equilibrium, which only very recently have been begun to be addressed~\cite{wang2015,koltai2016,pelle16}. Reducing the complexity by removing states implies a reduced entropy production, which severely influences dynamical properties and fluctuations~\cite{pugl10}. In Ref.~\citenum{knoch2015}, we have introduced non-equilibrium Markov state modeling (NE-MSM) performing the coarse-graining in cycle space instead of collecting configurations in metastable basins as for conventional MSM. The analog of these basins are now \emph{communities} of cycles with similar properties. The coarse-graining procedure preserves the entropy production of these communities, which makes our approach consistent with stochastic thermodynamics~\cite{seif12}.

\section{Model}

We study a single model polymer with $N=50$ beads in shear flow inspired by Ref.~\citenum{alex2006}. We employ Brownian dynamics (BD) simulations with 
\begin{equation}
  \dot\x_k = -\nabla_kU+\vec v(\x_k)+\boldsymbol\eta_k(t)
\end{equation}
for the bead positions $\x_k=(x_k,y_k,z_k)^T$, where $\vec v(\x)$ represents the shear flow. Interactions with solvent particles are modeled by a random force with correlations $\mean{\eta_k^\alpha(t)\eta_l^\beta(t')}=2\delta_{kl}\delta^{\alpha\beta}\delta(t-t')$, where upper indices label vector components. The potential energy $U=U_\text{nb}+U_\text{b}$ is split into the non-bonded short-ranged Lennard-Jones pair potentials $U_\text{nb}=\varepsilon_{\text{LJ}}\sum_{k<l}[r_{k,l}^{-12}-2r_{k,l}^{-6}]$ and bonds $U_\text{b}=\frac{\kappa}{2}\sum_{k=1}^{N-1}(r_{k+1,k}-1)^2$ that connect the nearest neighboring beads. Here, $r_{k,l}$ is the distance between the $k$-th and $l$-th bead, $\varepsilon_{\text{LJ}}=2.3$ determines the strength of the non-bonded potential, and $\kappa=100$ is the effective spring constant. All quantities have been non-dimensionalized by rescaling lengths with the bead diameter $\sig$ and timescales with the characteristic monomer diffusion time $\sig^2/(4D_0)$. Numerical values for the strain rate $\gd$ thus correspond to the Weissenberg number.

The polymer is driven into a non-equilibrium steady state through simple shear flow. While some scaling relations depend on hydrodynamic interactions~\cite{alex2006,wink06}, the qualitative behavior of the cyclic motion does not and in the following we neglect hydrodynamic interactions. As flow profile we choose 
\begin{equation}
\vec v(\x)=\gd(z-\zc)\vec e_x,
\end{equation}
where $\gd$ is the strain rate and $\zc$ is the $z$ component of the center of mass of the polymer. We found that this shift of the flow stabilizes the globular and stretched configurations as it increases the effective barrier for folding/unfolding and thus leads to a better separation between globular and extended states. Qualitatively, the same effect would be expected when including hydrodynamic interactions with the wall. Our conclusions do not depend on this detail. The polymer is grafted onto a repulsive planar surface (the $x$-$y$ plane with $z=0$) by fixing the position of the first bead to $\x_1=(0,0,0.5)$. Although simplified, this model reproduces the cyclic dynamics found in experiments~\cite{doyl00}.

\begin{figure}[t]
  \centering
  \includegraphics[width=\linewidth]{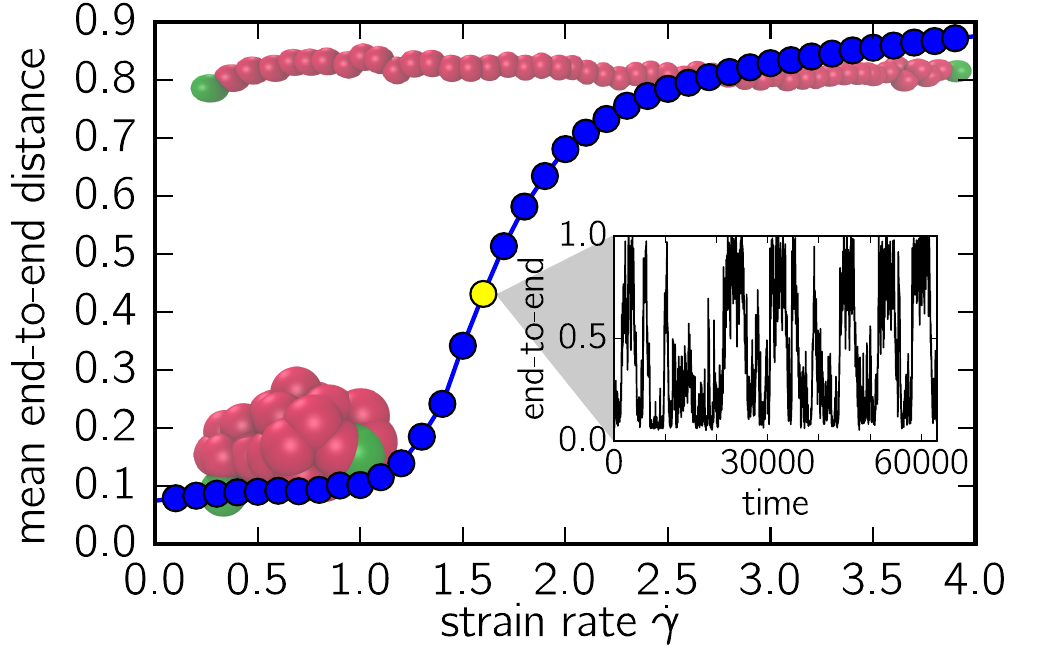}
  \caption{Mean end-to-end distance $\mean{\re}$ as a function of strain rate $\gd$. For $\gd=1.6$ (yellow/bright circle) the inset shows an exemplary time series. Also shown are two snapshots for the globule (bottom) and extended polymer (top).}
  \label{fig:polymer}
\end{figure}

A reasonable order parameter describing the folding and unfolding of the polymer is the relative end-to-end distance $\re\equiv|x_0-x_N|/N$, where $\re=1$ corresponds to a straight line of touching beads. We perform BD simulations for multiple values of $\dot{\gamma}$, see Fig.~\ref{fig:polymer}. We find different behaviors of $\mean{\re}$ that we categorize into three regimes. For $\gd\lesssim 1$ the polymer remains collapsed, while for $\gd\gtrsim2.2$ it is dominantly found in elongated conformations. For intermediate strain rates the polymer exhibits transitions between globular and elongated conformations, which was also found in similar simulations for free polymers~\cite{alex2006} and grafted polymers~\cite{lemak1998,delgado2006} under shear. The exemplary time series for $\gd=1.6$ in the inset of Fig.~\ref{fig:polymer} shows a clear separation of both states with random lifetimes and fast transitions. The average folding time $t_\text{fold}\approx3000$ is much larger than the intrinsic Rouse time $t_\text{R}=N^2/(2\pi^2\kappa)\sim1$ of the polymer.

\section{Theory}

Our goal is to construct a dynamically consistent NE-MSM with as few discrete states as possible. Every discrete state $i$ is a set of $N$ particle positions, which we collect in the vector $\X_i\equiv(\x_k)$ with $3N$ entries. These states, or \emph{centroids}, represent discrete volumes of configuration space, \emph{i.e.}, many configurations with slightly different positions. Following Ref.~\citenum{knoch2015}, the outline of the algorithm is as follows: The first step is to generate many (specifically $M=500$) centroids using the BD simulation data as input employing a spatial clustering algorithm (for technical details, see Ref.~\citenum{prinz2011} and appendix~\ref{sec:spatial}). From the counting matrix of transitions in the BD data we construct an approximate Markov process and identify cycles. The crucial step is to group cycles into \emph{communities} (\emph{i.e.}, clusters of cycles with similar properties defined through suitable order parameters) and determine one cycle representative for each community. In the actual coarse-graining, only representatives and the states that they visit are retained. The final step is then to rescale the Markov dynamics so that it preserves the entropy production of every community as well as the total entropy production.

Entropy production arises from non-vanishing probability currents $J^i_j\equiv\Phi^i_j-\Phi^j_i$ with probability fluxes $\Phi^i_j$ along transitions $i\to j$. In thermal equilibrium the condition of detailed balance holds, $\Phi^i_j=\Phi^j_i$, which implies zero currents. The mean entropy production rate~\cite{alta12a,knoch2015} is
\begin{equation}
  \label{sdot1}
  \dot S = \sum_{ij}\Phi^i_jA^i_j = \frac{1}{2}\sum_{ij}J^i_j A^i_j
\end{equation}
with \emph{affinities} $A^i_j\equiv\log{\Phi^i_j}-\log{\Phi^j_i}$.

\subsection{Cycle decomposition}

In a steady state, Kirchhoff's law implies that probability flows in cycles.  A cycle $\al=(i_1\to i_2\to\cdots\to i_1)$ is defined as an ordered set of states, at the end of which the starting state is reached again and all other states are visited exactly once. Cycles that differ only in the cyclic permutation of their states are considered identical. We extend the concept of affinities to cycles, yielding the cycle affinities $A_\al\equiv\sum_{(i\to j)\in \al} A^i_j$, where the sum is over all edges along cycle $\al$. Traversing a full cycle, the entropy produced equals the cycle affinity. Graph theoretical results allow us to decompose the probability fluxes~\cite{kalpazidou2007,knoch2015,alta12a}
\begin{equation}
  \label{decomp}
  \Phi^i_j = \sum_{\alpha\ni(i\to j)}\varphi_\alpha
\end{equation}
into cycle fluxes with non-negative weights $\varphi_\alpha\geqslant0$. The summation is over all cycles $\al$ that include the edge $(i\to j)\in\al$. The number of cycles quickly becomes very large. Here we employ the algorithm introduced in Ref.~\citenum{knoch2015} and summarized in appendix~\ref{sec:decomp} to efficiently determine a subset of cycles with non-vanishing weights $\varphi_\al>0$. These cycles have non-negative cycle affinities (and thus entropy production), which makes our approach conceptually different from the well-known Schnakenberg theory~\cite{schnakenberg1976}. Inserting Eq.~\eqref{decomp} into Eq.~\eqref{sdot1}, we obtain $\dot{S}=\sum_\alpha \varphi_\al A_\al$, which clearly shows that all entropy production is encoded in cycles.

\subsection{Cycle communities}

\begin{figure}[t]
  \centering
  \includegraphics[width=\linewidth]{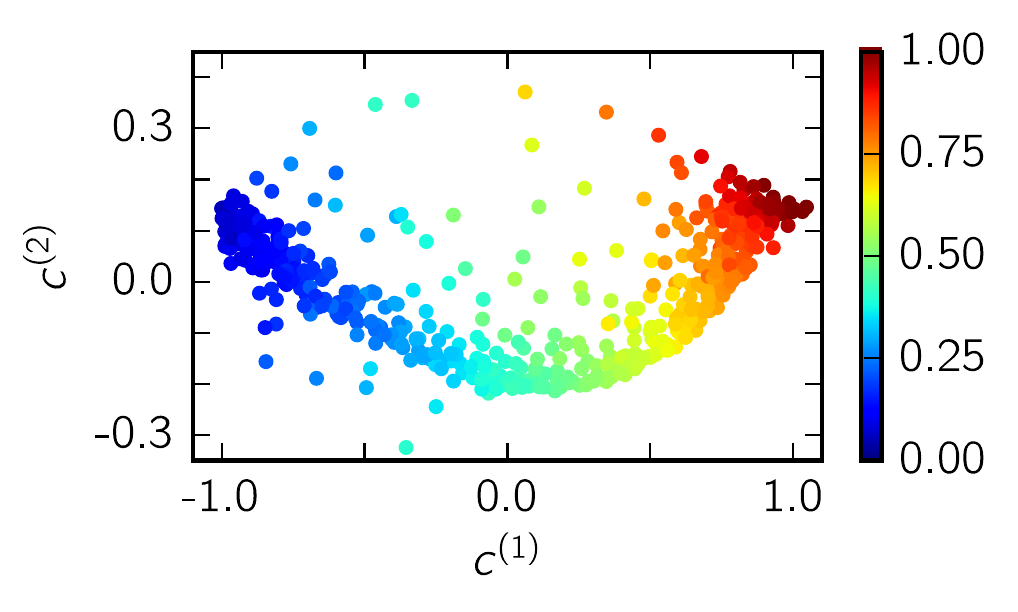}
  \caption{Configuration space ($\gd=1.6$). Scatter plot of the normalized projections of all centroids onto the two largest principal components. Colors indicate the end-to-end distance $\re$.}
  \label{fig:cnf}
\end{figure}

At this point we have determined all entropy-producing cycles. To make progress, we now need to find similarities between cycles. One approach is to consider the connectivity between cycles~\cite{alta12a,conrad2014modularity,knoch2015}. For the globule-stretch transition, we found this approach to be less suitable since it groups cycles that are located in different parts of configuration space~\footnote{The reason why the connectivity fails is presumably because the underlying potential energy landscape does not offer well separated basins.}. Instead, here we describe an alternative, general method to reduce the complexity of the high-dimensional configuration space to a few variables based on the topology of cycles. As our main tool, we perform a principal component analysis (PCA)~\cite{jolli2002}, which is an orthogonal linear transformation returning the eigenvectors (the principal components) and eigenvalues of the covariance matrix of the centroid positions. The principal component $\vec C^{(1)}$ corresponding to the largest eigenvalue coincides with the direction in configuration space exhibiting the largest variance; in the present case it already accounts for $\approx96$\% of the observed variance. In Fig.~\ref{fig:cnf} the normalized projections $c^{(n)}_i\equiv\X_i\cdot\vec C^{(n)}/\mathcal N$ of centroids onto first and second principal component are plotted, where the normalization $\mathcal N$ ensures that $c^{(1)}_i\in[-1,1]$. Centroids with small (negative) values for $c^{(1)}_i$ correspond to globular configurations (they contribute little to the observed variance of positions), large (positive) values correspond to stretched configurations. The second component indicates the variance within these states. Both globular and stretched configurations show larger fluctuations while the intermediate states with $c^{(1)}_i\sim 0$ exhibit less fluctuations. Hence, the PCA reproduces the expected, typical picture of two basins with intermediate transition states.

\begin{figure}
  \centering
  \includegraphics[width=\linewidth]{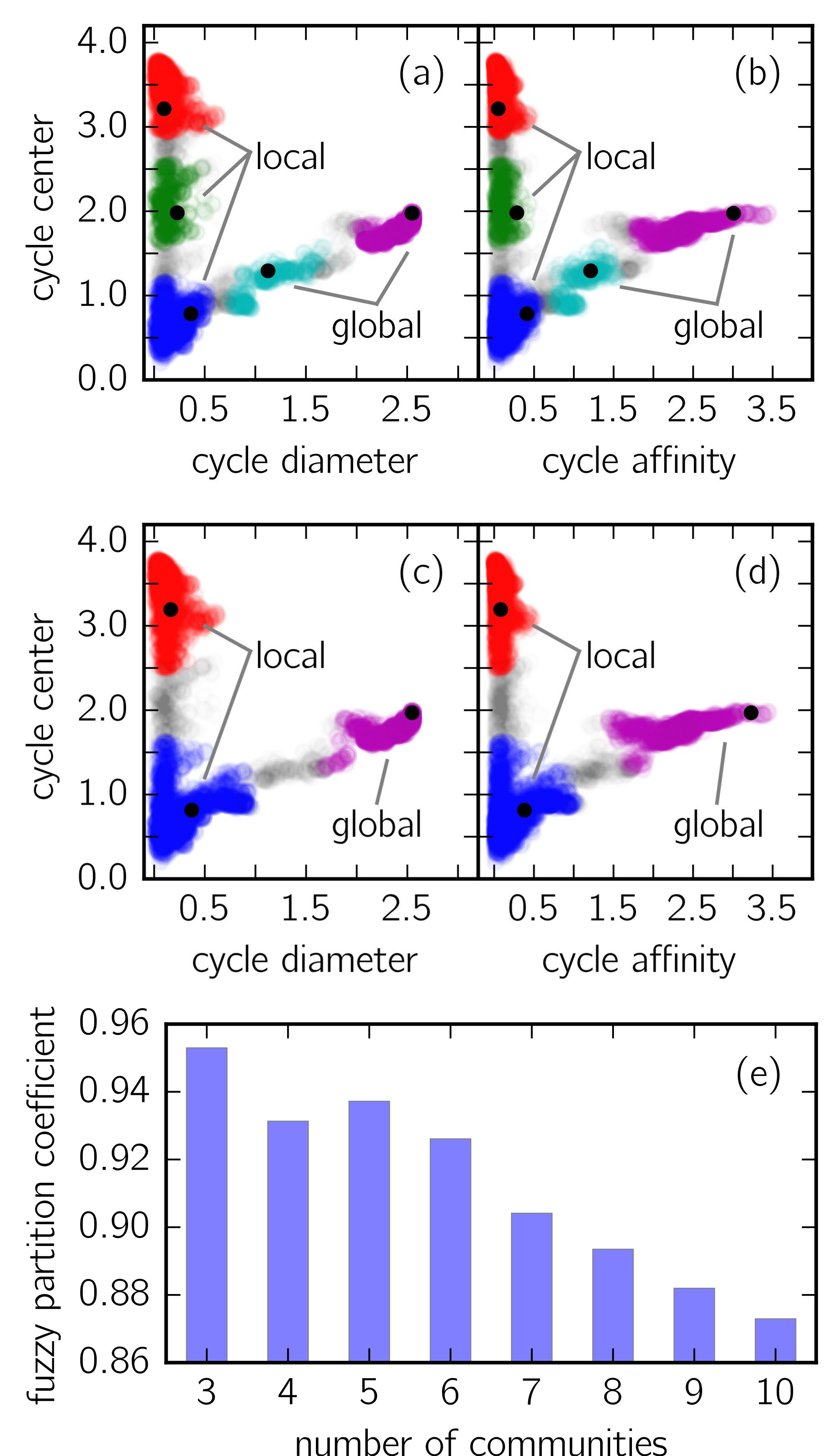}
  \caption{Cycle space ($\gd=1.6$). (a)~Scatter plot of cycle center vs. the cycle diameter, and (b)~cycle affinity vs. cycle center. The cycle diameter and centers are computed using a PCA (see text for details). Colors represent cycle communities while gray points indicate no community. The filled black circles indicate the cycle representatives. (c,d)~The same plots as for (a,b) but for three cycle communities. (e)~The fuzzy partition coefficients computed for multiple communities. The best result is obtained for 3 communities followed by a slightly lower value for 5 communities.}
  \label{fig:cyc}
\end{figure}

Going back to cycles, we now define two variables: (i)~the cycle centers
\begin{equation}
  x_\al \equiv \left(\frac{1}{|\al|}\sum_{i\in\al}\X_i\right)\cdot\vec C^{(1)}
\end{equation}
with $|\al|$ the number of centroids in cycle $\al$ and (ii)~the cycle diameters
\begin{equation}
  d_\al \equiv \max_{i,j\in\al}\left\{ |c^{(1)}_i-c^{(1)}_j| \right\}.
\end{equation}
The set of points $(x_\al,d_\al)$ is plotted in Fig.~\ref{fig:cyc}a, where every point now represents a single cycle. These points are clearly not random. Many cycles have a small diameter $d_\al$ but different cycle centers. We identify these cycles as \emph{local} because the centroids $i\in\al$ in these cycles have similar $c^{(1)}_i$ and thus belong to a compact region in configuration space. There is a second group of cycles with large diameter, which, consequently, we identify as \emph{global} cycles. Basically the same structure is recovered when plotting $(x_\al,A_\al)$ shown in Fig.~\ref{fig:cyc}b, indicating that local cycles have low affinities and global cycles have large affinities.

We can turn these insights into a more quantitative statement by partitioning the cycle space into $k$ communities. To this end we employ an implementation of the fuzzy $c$-means algorithm (an implementation is available in Ref.~\citenum{cmeans}), which assigns to each cycle a probability for belonging to a specific community. First, we normalize all three features (cycle diameters, cycle centers, and affinities) by their variance to make them comparable. These features are then used as an input for the fuzzy c-means clustering algorithm returning membership degrees $u_{ij}$ which express the probability that observation $i$ belongs to community $j$. To obtain an indicator of how good the clustering results are we compute the fuzzy partition coefficient (FPC) that is defined as the Frobenius norm of the membership matrix 
\begin{equation}
 \text{FPC} = \frac{1}{n}\sum_{i=1}^k\sum_{j=1}^n u_{ij}^2.
\end{equation}
Here $k$ is the number of chosen communities and $n$ the number of observations (cycles in our case). The closer the FPC gets to one the better the cycle space can be partitioned into the chosen number of clusters.

The advantage of fuzzy partitioning is that some cycles might not belong to any cycle community while others match well in multiple ones. The minimal number of cycle communities to account for the collective folding and unfolding dynamics is $k=3$ (see Fig.~\ref{fig:cyc}e), but to capture the full dynamics we found $k=5$ cycle communities to be more appropriate. We identify three local communities colored in blue, green, and red, as well as two global communities colored in magenta and cyan, see Fig.~\ref{fig:cyc}a,b. The blue community represents cycles that correspond to globular configurations while the green and red communities represent similar dynamics for intermediate and fully stretched configurations, respectively. The global cycle communities connect two (cyan) or all three (magenta) local communities.

\subsection{Coarse-graining}

Next, we replace each cycle community by one cycle that we will refer to as \emph{representative}~\cite{knoch2015}. We find appropriate representatives by mapping mean first-passage times between states that belong to different local cycle communities. After selecting representatives, we delete all states not belonging to any of the representatives and rescale the remaining transition rates with restrictions that: the total entropy production rate $\dot{S}$, all remaining cycle affinities $A_\alpha$ and all remaining edge affinities $A^i_j$ are preserved. For details see Ref.~\citenum{knoch2015} and appendix~\ref{sec:repr}.


At this stage the coarse-grained MSM still contains many states since a single cycle can traverse hundreds of states. The important point is, however, that the coarse-grained model lost much of its original complexity as it now contains only a few cycles. We can thus further reduce the number of states. To this end we identify two dominant motifs, which we refer to as bridge and triangle states. Both motifs build on states that have exactly two neighbors. For bridge states the neighbors are not connected to each other~\cite{alta2012}. Triangle states complete cycles that do not exhibit positive entropy productions and thus are not part of the decomposition Eq.~\eqref{decomp}. We iteratively search and remove these states (for details see appendix~\ref{sec:bridge}) until no more are found, which completes our coarse-graining scheme.

\section{Coarse-grained NE-MSM}

\begin{figure}
  \centering
  \includegraphics[width=\linewidth]{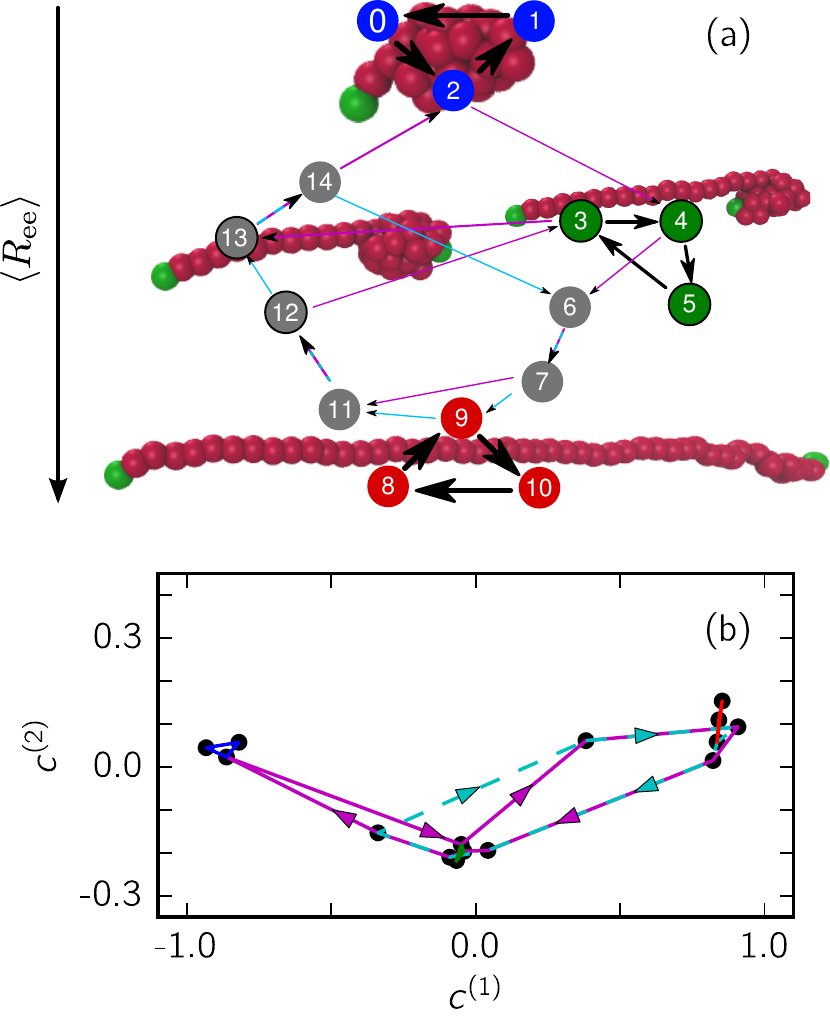}
  \caption{Final NE-MSM for $\gd=1.6$. (a)~Transition network of the polymer dynamics with 15 centroids (filled circles) and 5 cycles. The colored states green, blue, and red correspond to the colors of the local cycle communities shown in Fig.~\ref{fig:cyc}. States with a black border are structurally very similar and constitute the transition ensemble. The arrows point in the direction of probability currents (net flux), while the arrow widths represent the magnitude of currents. On average, the polymer dynamics follows the direction of the arrows. Arrow colors cyan and magenta correspond to the global cycles. (b)~Transition network in configuration space using the normalized projections onto the two largest principal components. Symbols are the configurations while lines indicate the cycles with the arrows pointing in the direction of the net flux.}
  \label{fig:tnetwork}
\end{figure}

The final NE-MSM for strain rate $\gd=1.6$ is shown in Fig.~\ref{fig:tnetwork}a. After removing the bridge and triangle states the transition network contains 15 states (centroids). The collective rates for folding and unfolding of this coarse-grained model agree with those obtained from the BD simulations by construction. Moreover, the remaining five cycles now allow for detailed insights into the relevant pathways in non-equilibrium. The three local cycles are composed of three states, the minimal number for a non-trivial, entropy-producing cycle. The red and blue cycle connect stable globular and stretched configurations, respectively. The green cycle represents a metastable intermediate of half-stretched configurations that do not unfold correctly but quickly fold back to the intermediate. The global cyan cycle also contains half-stretched configurations (structurally similar to the green cycle) but here the unfolding reaches the final stretched states before returning to their half-stretched origin. Finally, the magenta cycle represents the full transition from globule to stretched configurations and back.

The five states $(3,4,5,12,13)$ describing intermediate, half-stretched configurations are very close in configuration space (see Fig.~\ref{fig:tnetwork}b). They constitute the analog of the transition ensemble through which the folding/unfolding has to proceed. In non-equilibrium, however, the folding and unfolding processes follow different paths through this narrow region of configuration space. The globule to half-stretched transition proceeds along $2\to4$ (with state 4 belonging to the green cycle) whereas the reverse half-stretched to globule transition proceeds along $13\to14\to2$, with states 13 and 14 belonging to the cyan cycle. The cycle topology thus reveals the dynamical trapping of the polymer in an intermediate, which cannot be captured by structural information alone. Employing order parameters like distances in configuration space or lifetimes to identify mesostates will clearly miss this important feature of cyclic non-equilibrium dynamics.

Another question that we can address is dissipation, the role of which for bio-processes has been investigated recently, \emph{e.g.}, for self-replication~\cite{engl13} and in the activation of signaling proteins~\cite{weber2015}. The rate of dissipated heat $\dot Q_\text{diss}=\dot S/T$ created in each cycle community is proportional to their respective entropy production rate $\dot S$. Our analysis reveals that both the blue and red cycles are equally responsible for about 30\% of the total dissipated heat, while the green, cyan, and magenta communities produce 5\%, 15\%, and 50\%, respectively. The latter is caused by the large conformational changes (folding and unfolding process) of the polymer. The blue and red cycles, on the other hand, do not exhibit large conformational changes, therefore, the conformational changes must be on very short timescales, which is confirmed by the large probability currents shown in Fig.~\ref{fig:tnetwork}a.

We finally comment on using a PCA, which is not necessarily the best method to reduce the dimensionality of configuration space, and more advanced methods such as the time-lagged independent correlation analysis exist~\cite{perez2013}. Moreover, it is common practice to first apply a dimensionality reduction to the continuous molecular dynamics (or BD) data and subsequently discretize the reduced configuration space into finite sets. We did not follow this approach because NE-MSMs built from reduced configuration space exhibit a significantly lower entropy production rate than NE-MSMs built from full configuration space (assuming the same number of centroids is used). Recent results~\cite{bara15,ging16} suggest that the fluctuations of general currents are bounded by the entropy production, which directly links dynamic with thermodynamic consistency.


\section{Conclusions}

To conclude, we have presented a general method to systematically construct coarse-grained models composed of a few discrete states from molecular data of steadily driven systems. Specifically, we have studied the globule-stretch transition of a simple model polymer, but the method can in principle be applied to more realistic and complex molecules (such as F$_1$-ATPase~\cite{sugawa2016}), delivering minimal and thermodynamically consistent Markov state models. While these models can be employed to bridge time scales, they also allow insights into the relevant transition pathways. Here we have shown that different directions can pass through the same region of configuration space, which we believe might be a general property of transitions in driven systems.


\acknowledgments

We thank F. Schmid for helpful comments. We acknowledge financial support by the DFG through the collaborative research center TRR 146 (project A7).

\appendix

\section{Spatial clustering}
\label{sec:spatial}

The configuration space is discretized by employing the popular $k$-means clustering algorithm which divides the continuous configuration space into $k$ volumes and returns their centroids (center of each volume).
Next the BD trajectories are projected onto the centroids, storing its dynamical information as simple sequences of centroid indices.
We extract the dynamical information by counting the number of transitions $C_{i\to j}(\tau)$ for a given lag time $\tau$ (time in between transitions). The count matrix $\mat C(\tau)$ is used to approximate the transition matrix $\mat T(\tau)$ by
\begin{equation}
  T_{i\to j}(\tau) = \frac{C_{i\to j}(\tau)}{\sum_j C_{i\to j}(\tau)},
\end{equation}
which is also the maximum probability estimator for the true transition operator~\cite{prinz2011}. We additionally require $\mat T$ to have two important properties: First, $\mat T$ must be ergodic, i.e. each state can be reached from every other state in finite time or the transition network spanned by $\mat T$ is connected. Second, all occurring transitions are reversible, i.e. if $T^i_j>0$ then $T^j_i>0$.

\begin{figure}[b]
 \centering
 \includegraphics[width=\linewidth]{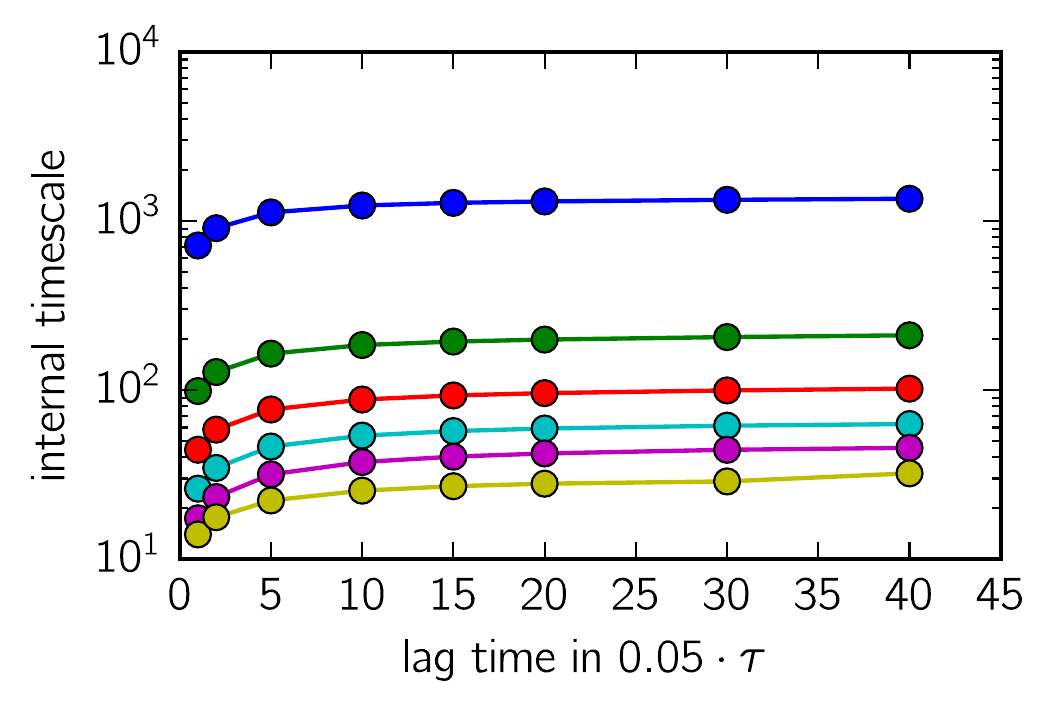}
  \caption{Lag time analysis. The six slowest internal timescales are plotted for different lag times. For building the MSM we choose $\tau=20$.}
 \label{lagtime_analysis}
\end{figure}

If the process described by $\mat T$ is time-homogeneous its entries $T^i_j$ are ought to be time-independent. To test this we perform a lag time analysis~\cite{swo2004} shown in Fig.~\ref{lagtime_analysis}. For equilibrium systems the eigenvalues $\lambda_i$ of the transition matrix $\mat T$ can be converted into relaxation timescales of the system $t_i = -\tau/\log{(\lambda_i)}$ with slowest modes corresponding to the largest $\lambda_i$. If the largest eigenvalues are time-independent, also the slowest modes are time-independent yielding a good estimation of the minimal lag time that can be used. In NE-MSM the eigenvalues can become complex, which leaves their physical interpretation still open for discussion.
However, for the polymer dynamics the largest eigenvalues sorted by either their real part or absolute value exhibit no imaginary part.

For a continuous-time Markov process with rate matrix $\mat W$, the transition probability can be expressed as $\mat T=e^{\mat W\tau}$. While conventional Markov state models have discrete-time dynamics, for our purposes we require a rate matrix, which we approximate from the computed transition matrix through
\begin{equation}
  \mat W \approx \frac{\mat T - \mathbf{1}}{\tau}.
  \label{eq:transition_matrix}
\end{equation}
All off-diagonal elements of $\mat W$ are non-negative $\omega^i_j\ge 0$ while the diagonal elements are $\omega^i_i = -\sum_j \omega^i_j$. Following the Perron-Frobenius theorem, $\mat T$ and thus $\mat W$ have an unique largest real eigenvalue $\lambda_\text{max}$ with a corresponding eigenvector that has strictly positive entries. This eigenvector represents the steady-state probability distribution with elements $p_i$. Finally, the probability fluxes are obtained by $\Phi^i_j = \omega^i_j p_i$.

\section{Cycle decomposition}
\label{sec:decomp}

Here we describe an algorithm that decomposes a given probability flux matrix $\mat \Phi$ into cycle fluxes. Each element of $\mat \Phi$ transforms through $\Phi^i_j = \sum_{(i\to j)\ni \alpha}\varphi_\alpha$ with  $\varphi_\alpha \ge 0$. The sum adds the cycle weights of all cycles containing the edge $i\to j$.
For the cycle decomposition we start by dividing $\mat \Phi$ into a symmetric detailed-balance part $\mat \Phi^{\text{db}}$ and a non-negative current part $\mat J$ so that 
\begin{equation}
 \mat \Phi = \mat \Phi^{\text{db}} + \mat J.
\end{equation}
Here $\mat J$ is obtained by $\mat \Phi - \mat \Phi^T$ with all negative elements set to zero. The symmetric part follows as $\mat \Phi^{\text{db}} = \mat \Phi - \mat J$. To make progress, we identify all non-zero elements of $\mat \Phi^{\text{db}}$ yielding a set of trivial cycles, i.e. cycles with only two different states ($i\to j \to i$). Their cycle weights are identical to their corresponding non-zero entry $\varphi_{i\to j\to i} \equiv \Phi^{i\,\text{db}}_j = \Phi^{j\,\text{db}}_i$.
The algorithm to decompose the current part $\mat J$ is split into two parts. First, searching for a non-trivial cycles (cycle with more than 2 different states) and second determining its cycle weights.
Theoretically, the number of possible non-trivial cycles grows exponentially with the number of non-zero entries of $\mat J$. However, if both steps run alternately the decomposition becomes computationally affordable even for a large number of states.
To detect a non-trivial cycle, we propose the following steps:
\begin{enumerate}
 \item Find the position of the largest element of $\mat J$, $\operatorname {arg\,max}(J^i_j)$.
 \item Search for the shortest path (smallest number of transitions) from state $j$ leading back to state $i$ (only following the non-zero transitions). This step can be efficiently achieved by applying a breadth-fist/depth-first search~\cite{newman2010}.
 \item Return the non-trivial cycle, i.e. ($i\to j \to\, $found path).
\end{enumerate}
To determine the corresponding cycle weight $\varphi_\alpha$, we take all flux values along cycle $\alpha$ and determine their smallest value becoming the cycle weight
\begin{equation}
  \varphi_\alpha \equiv \min_{i\to j\in\alpha}\{J^i_j\}.
\end{equation}   
Summing up both steps, the final algorithm reads
\begin{enumerate}
 \item Find a non-trivial cycle
 \item Compute its cycle weight $\varphi_\alpha$
 \item Update the current matrix by subtracting $\varphi_\alpha$ along $\alpha$,  $\mat J = \mat J - \sum_{(i\to j)\in\alpha} \varphi_\alpha$ and repeat with the first step
 \item The algorithm stops when the residuum $||\mat J_{\text{max}}||$ has become smaller than a threshold.
\end{enumerate}
General considerations~\cite{kalpazidou2007,alta12a} show that the maximum number of needed non-trivial cycles is bounded by $N_\text{cycles} =|E|-|V|+1$ with $|E|$ being the number of non-zero elements of $\mat J$ and $|V|$ its rank.

\section{Cycle representatives and coarse-graining}
\label{sec:repr}

Once the cycle communities are found, the idea for coarse-graining the MSM is to pick one cycle for each community -- we refer to it as representative --, delete all states not belonging to any representative and, finally, rescale the remaining transition rates $\omega^i_j$. To be thermodynamically consistent the newly computed transition rates have to preserve four quantities: The total entropy production rate $\dot S$, the cycle affinities of the representatives $A_\alpha$, all remaining edge affinities $A^i_j$ and the dissipated heat along the remaining edges $\log{(\omega^i_j)} - \log{(\omega^j_i)}$. For a detailed description of the coarse-graining algorithm we refer the reader to our previous publication~\cite{knoch2015}.

Since we know now how to coarse-grain a given set of cycle representatives, we address the question of how to select ``appropriate'' representatives.
Any set of cycle representatives is thought to be appropriate if the graph spanned by their coarse-grained transition matrix is ergodic and the mean first passage times (MFPT) between local communities are preserved. For example, the polymer dynamics for $k=3$ communities as illustrated in Fig.~\ref{fig:cyc}c,d exhibits 2 local and 1 global community. To compute the MFPTs we identify all states of the red community, say as set $R$, and all states belonging to the blue one as set $B$. Any appropriate set of representatives needs to preserve MFPT$_{R\to B}$ and MFPT$_{B\to R}$.

Especially, the conservation of MFPTs is of particular importance as it ensures the coarse-grained MSM to express the correct timescales. So far no determinstic algorithm exists that returns a reliable set of representatives. For this reason we formulate a stochastic algorithm that picks candidates for cycle representatives randomly and checks for ergodicity and MFPTs. The algorithm is outlined as follows:
\begin{enumerate}
\item Choose one representative per cycle community by drawing a random number.
\item Check if set of representatives span ergodic transition network. If yes, compute coarse-grained MSM, else go back to step (1).
\item Compute MFPTs of the coarse-grained MSM and compare to MFPTs of full MSM. If MFPTs match, return coarse-grained MSM, else go back to step (1).
\end{enumerate}

\section{Coarse-graining of bridge and triangle states}
\label{sec:bridge}

\begin{figure}[b!]
  \centering
  \includegraphics[width=0.75\linewidth]{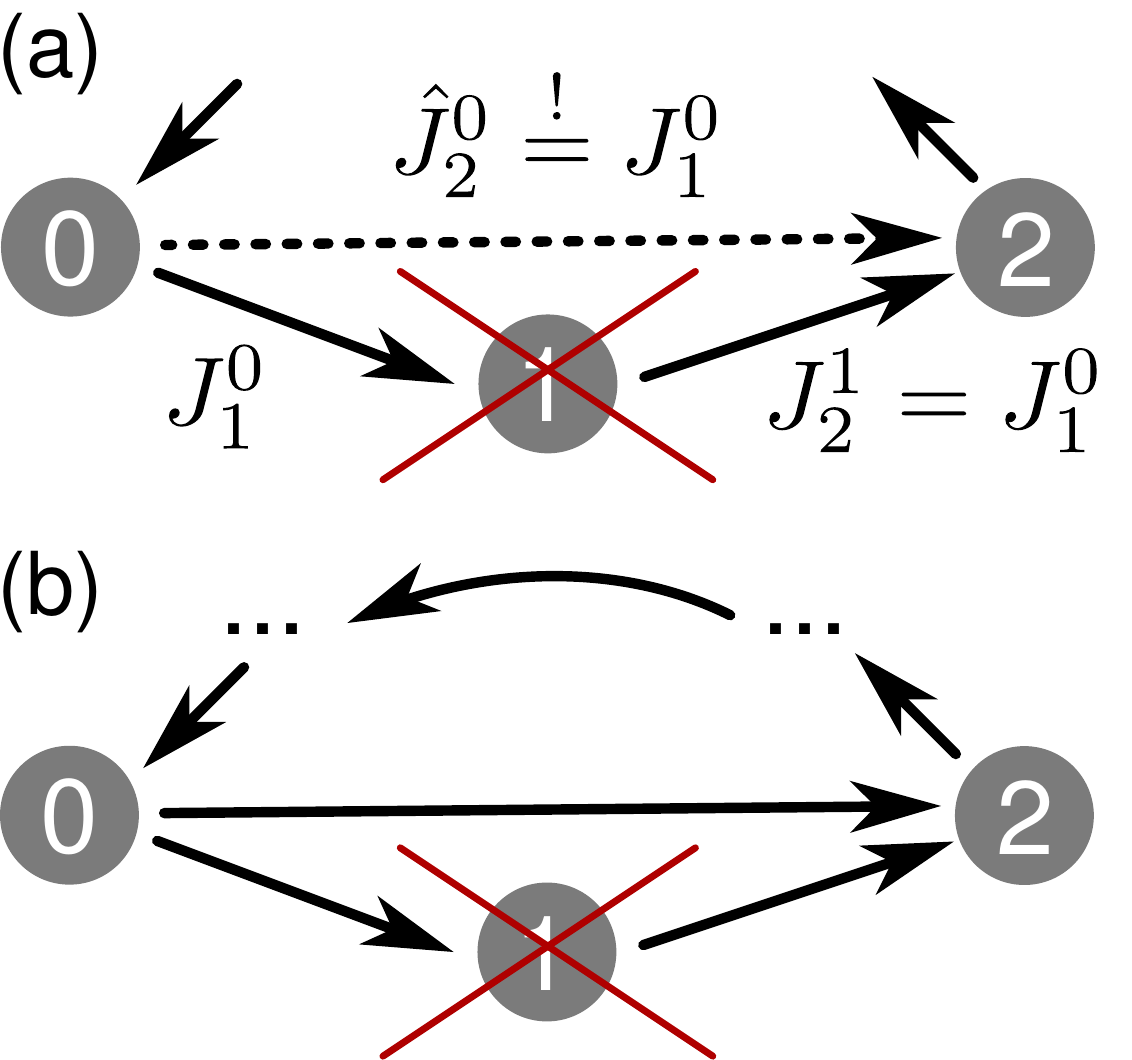}
  \caption{Illustration of bridge and triangle state coarse-graining approach. (a) Bridge state: After deletion of state 1 the other states are directly connected (dotted arrow). (b) Triangle state: After deletion of state 1 the existing connections $0\leftrightarrow2$ are modified. The arrows point in the direction of probability currents.}
  \label{bridgestate}
\end{figure}

To coarse-grain bridge states (state 1 in Fig.~\ref{bridgestate}a) we form a new connection between the two neighboring states (state 0 and 2 in Fig.~\ref{bridgestate}a). The new transition rates $\hat \omega^0_2$ and $\hat \omega^2_0$ have to preserve three characteristics: 
(i) edge affinities $\hat{A}^0_2 = A^0_1 + A^1_2$, (ii) probability currents $\hat{J}^0_2 = J^0_1 = J^1_2 $ and (iii) dissipated heat $\log{(\hat\omega^0_2/\hat\omega^2_0)} = \log{(\omega^0_1/\omega^1_0)} +\log{(\omega^1_2/\omega^2_1)}$.
Using conditions (i) and (ii), the new probability fluxes follow as
\begin{align}
\nonumber
 \hat\Phi^0_2 &= \hat\Phi^2_0\frac{\Phi^0_1\Phi^1_2}{\Phi^1_0\Phi^2_1}~\text{ and }~\hat\Phi^0_2 = \hat\Phi^2_0 + \Phi^0_1 - \Phi^1_0\\
 \Rightarrow \hat\Phi^2_0 &= \frac{\Phi^0_1 - \Phi^1_0}{\left(\frac{\Phi^0_1\Phi^1_2}{\Phi^1_0\Phi^2_1}\right) - 1}
\end{align}
To fulfill condition (iii) we demand that the ratio between any two probabilities of the full network are preserved $\hat p_i/\hat p_j = p_i/p_j$, which includes $\hat p_0/\hat p_2 = p_0/p_2$, and thus
\begin{align}
 \frac{w^i_j}{w^j_i} = \frac{\Phi^{i}_{j}}{\Phi^{j}_{i}}
  \frac{p_j}{p_i} = \frac{\hat{\Phi}^{i}_{j}}{\hat{\Phi}^{j}_{i}}
  \frac{\hat{p}_j}{\hat{p}_i} = \frac{\hat{w}^{i}_{j}}{\hat{w}^{j}_{i}}.
\end{align}
Condition (iii) is also the main difference between our approach and the one discussed in Ref.~\citenum{alta2012}. In our adaptation the probability distribution of the complete network is changed, while Altaner \ea change it only locally ($p_0$ and $p_2$) and hence absorb $p_1$ into $p_0$ and $p_2$. The disadvantage of the latter is that condition (iii) is not preserved for the full network and, when using the coarse-graining approach iteratively, accumulation of probability in single states might occur, which leads to unphysical results.

For the coarse-graining of triangle states (state 1 in Fig.~\ref{bridgestate}b) we consider all cycles that contain the edges $0\to1\to2$. We modify these cycles by replacing the edges $0\to1\to2$ with a new edge $0\to2$. To be thermodynamically consistent the modified cycles have to (i) preserve the cycle entropy production rate $\dot S_\alpha \equiv \varphi_\alpha A_\alpha$, (ii) preserve the edge affinities $A^0_2$ and (iii) dissipated heat $\log{(\omega^0_2/\omega^2_0)}$. To rescale the transition rates we use the same rescaling algorithm as for the rescaling of the cycle representatives. Note that through restriction (i) it is not necessarily possible to coarse-grain all found triangle structures. Assume, for instance, that the modified cycle coincides with an already existing cycle, then, the rescaling is not unique anymore and entropy production is destroyed as only one out of two cycles survives.


%


\begin{thebibliography}{42}%
\makeatletter
\providecommand \@ifxundefined [1]{%
 \@ifx{#1\undefined}
}%
\providecommand \@ifnum [1]{%
 \ifnum #1\expandafter \@firstoftwo
 \else \expandafter \@secondoftwo
 \fi
}%
\providecommand \@ifx [1]{%
 \ifx #1\expandafter \@firstoftwo
 \else \expandafter \@secondoftwo
 \fi
}%
\providecommand \natexlab [1]{#1}%
\providecommand \enquote  [1]{``#1''}%
\providecommand \bibnamefont  [1]{#1}%
\providecommand \bibfnamefont [1]{#1}%
\providecommand \citenamefont [1]{#1}%
\providecommand \href@noop [0]{\@secondoftwo}%
\providecommand \href [0]{\begingroup \@sanitize@url \@href}%
\providecommand \@href[1]{\@@startlink{#1}\@@href}%
\providecommand \@@href[1]{\endgroup#1\@@endlink}%
\providecommand \@sanitize@url [0]{\catcode `\\12\catcode `\$12\catcode
  `\&12\catcode `\#12\catcode `\^12\catcode `\_12\catcode `\%12\relax}%
\providecommand \@@startlink[1]{}%
\providecommand \@@endlink[0]{}%
\providecommand \url  [0]{\begingroup\@sanitize@url \@url }%
\providecommand \@url [1]{\endgroup\@href {#1}{\urlprefix }}%
\providecommand \urlprefix  [0]{URL }%
\providecommand \Eprint [0]{\href }%
\providecommand \doibase [0]{http://dx.doi.org/}%
\providecommand \selectlanguage [0]{\@gobble}%
\providecommand \bibinfo  [0]{\@secondoftwo}%
\providecommand \bibfield  [0]{\@secondoftwo}%
\providecommand \translation [1]{[#1]}%
\providecommand \BibitemOpen [0]{}%
\providecommand \bibitemStop [0]{}%
\providecommand \bibitemNoStop [0]{.\EOS\space}%
\providecommand \EOS [0]{\spacefactor3000\relax}%
\providecommand \BibitemShut  [1]{\csname bibitem#1\endcsname}%
\let\auto@bib@innerbib\@empty
\bibitem [{\citenamefont {Karplus}\ and\ \citenamefont
  {McCammon}(2002)}]{karp02}%
  \BibitemOpen
  \bibfield  {author} {\bibinfo {author} {\bibfnamefont {M.}~\bibnamefont
  {Karplus}}\ and\ \bibinfo {author} {\bibfnamefont {J.~A.}\ \bibnamefont
  {McCammon}},\ }\bibfield  {title} {\enquote {\bibinfo {title} {Molecular
  dynamics simulations of biomolecules},}\ }\href {\doibase
  10.1038/nsb0902-646} {\bibfield  {journal} {\bibinfo  {journal} {Nat. Struct.
  Mol. Biol.}\ }\textbf {\bibinfo {volume} {9}},\ \bibinfo {pages} {646--652}
  (\bibinfo {year} {2002})}\BibitemShut {NoStop}%
\bibitem [{\citenamefont {Kamerlin}\ \emph {et~al.}(2011)\citenamefont
  {Kamerlin}, \citenamefont {Dryga},\ and\ \citenamefont {Warshel}}]{kame11}%
  \BibitemOpen
  \bibfield  {author} {\bibinfo {author} {\bibfnamefont {S.~C.~L.}\
  \bibnamefont {Kamerlin}}, \bibinfo {author} {\bibfnamefont {S.~V.~A.}\
  \bibnamefont {Dryga}}, \ and\ \bibinfo {author} {\bibfnamefont
  {A.}~\bibnamefont {Warshel}},\ }\bibfield  {title} {\enquote {\bibinfo
  {title} {Coarse-grained (multiscale) simulations in studies of biophysical
  and chemical systems},}\ }\href {\doibase
  10.1146/annurev-physchem-032210-103335} {\bibfield  {journal} {\bibinfo
  {journal} {Ann. Rev. Phys. Chem.}\ }\textbf {\bibinfo {volume} {62}},\
  \bibinfo {pages} {41--64} (\bibinfo {year} {2011})}\BibitemShut {NoStop}%
\bibitem [{\citenamefont {M\"uller-Plathe}(2002)}]{mull02}%
  \BibitemOpen
  \bibfield  {author} {\bibinfo {author} {\bibfnamefont {F.}~\bibnamefont
  {M\"uller-Plathe}},\ }\bibfield  {title} {\enquote {\bibinfo {title}
  {Coarse-graining in polymer simulation: From the atomistic to the mesoscopic
  scale and back},}\ }\href@noop {} {\bibfield  {journal} {\bibinfo  {journal}
  {ChemPhysChem}\ }\textbf {\bibinfo {volume} {3}},\ \bibinfo {pages}
  {754--769} (\bibinfo {year} {2002})}\BibitemShut {NoStop}%
\bibitem [{\citenamefont {Harmandaris}\ and\ \citenamefont
  {Kremer}(2009)}]{harm09}%
  \BibitemOpen
  \bibfield  {author} {\bibinfo {author} {\bibfnamefont {V.~A.}\ \bibnamefont
  {Harmandaris}}\ and\ \bibinfo {author} {\bibfnamefont {K.}~\bibnamefont
  {Kremer}},\ }\bibfield  {title} {\enquote {\bibinfo {title} {Dynamics of
  polystyrene melts through hierarchical multiscale simulations},}\ }\href
  {\doibase 10.1021/ma8018624} {\bibfield  {journal} {\bibinfo  {journal}
  {Macromol.}\ }\textbf {\bibinfo {volume} {42}},\ \bibinfo {pages} {791--802}
  (\bibinfo {year} {2009})}\BibitemShut {NoStop}%
\bibitem [{\citenamefont {Salerno}\ \emph {et~al.}(2016)\citenamefont
  {Salerno}, \citenamefont {Agrawal}, \citenamefont {Perahia},\ and\
  \citenamefont {Grest}}]{sale16}%
  \BibitemOpen
  \bibfield  {author} {\bibinfo {author} {\bibfnamefont {K.~M.}\ \bibnamefont
  {Salerno}}, \bibinfo {author} {\bibfnamefont {A.}~\bibnamefont {Agrawal}},
  \bibinfo {author} {\bibfnamefont {D.}~\bibnamefont {Perahia}}, \ and\
  \bibinfo {author} {\bibfnamefont {G.~S.}\ \bibnamefont {Grest}},\ }\bibfield
  {title} {\enquote {\bibinfo {title} {Resolving dynamic properties of polymers
  through coarse-grained computational studies},}\ }\href {\doibase
  10.1103/PhysRevLett.116.058302} {\bibfield  {journal} {\bibinfo  {journal}
  {Phys. Rev. Lett.}\ }\textbf {\bibinfo {volume} {116}},\ \bibinfo {pages}
  {058302} (\bibinfo {year} {2016})}\BibitemShut {NoStop}%
\bibitem [{\citenamefont {No{\'e}}\ and\ \citenamefont
  {Fischer}(2008)}]{noe2008}%
  \BibitemOpen
  \bibfield  {author} {\bibinfo {author} {\bibfnamefont {F.}~\bibnamefont
  {No{\'e}}}\ and\ \bibinfo {author} {\bibfnamefont {S.}~\bibnamefont
  {Fischer}},\ }\bibfield  {title} {\enquote {\bibinfo {title} {Transition
  networks for modeling the kinetics of conformational change in
  macromolecules},}\ }\href@noop {} {\bibfield  {journal} {\bibinfo  {journal}
  {Curr. Opin. Struct. Biol.}\ }\textbf {\bibinfo {volume} {18}},\ \bibinfo
  {pages} {154--162} (\bibinfo {year} {2008})}\BibitemShut {NoStop}%
\bibitem [{\citenamefont {No{\'e}}\ \emph {et~al.}(2009)\citenamefont
  {No{\'e}}, \citenamefont {Sch{\"u}tte}, \citenamefont {Vanden-Eijnden},
  \citenamefont {Reich},\ and\ \citenamefont {Weikl}}]{noe2009}%
  \BibitemOpen
  \bibfield  {author} {\bibinfo {author} {\bibfnamefont {F.}~\bibnamefont
  {No{\'e}}}, \bibinfo {author} {\bibfnamefont {C.}~\bibnamefont
  {Sch{\"u}tte}}, \bibinfo {author} {\bibfnamefont {E.}~\bibnamefont
  {Vanden-Eijnden}}, \bibinfo {author} {\bibfnamefont {L.}~\bibnamefont
  {Reich}}, \ and\ \bibinfo {author} {\bibfnamefont {T.R.}\ \bibnamefont
  {Weikl}},\ }\bibfield  {title} {\enquote {\bibinfo {title} {Constructing the
  equilibrium ensemble of folding pathways from short off-equilibrium
  simulations},}\ }\href@noop {} {\bibfield  {journal} {\bibinfo  {journal}
  {Proc. Natl. Acad. Sci. USA}\ }\textbf {\bibinfo {volume} {106}},\ \bibinfo
  {pages} {19011--19016} (\bibinfo {year} {2009})}\BibitemShut {NoStop}%
\bibitem [{\citenamefont {Bowman}\ \emph {et~al.}(2011)\citenamefont {Bowman},
  \citenamefont {Voelz},\ and\ \citenamefont {Pande}}]{bowm11}%
  \BibitemOpen
  \bibfield  {author} {\bibinfo {author} {\bibfnamefont {G.~R.}\ \bibnamefont
  {Bowman}}, \bibinfo {author} {\bibfnamefont {V.~A.}\ \bibnamefont {Voelz}}, \
  and\ \bibinfo {author} {\bibfnamefont {V.~S.}\ \bibnamefont {Pande}},\
  }\bibfield  {title} {\enquote {\bibinfo {title} {Taming the complexity of
  protein folding},}\ }\href {\doibase 10.1016/j.sbi.2010.10.006} {\bibfield
  {journal} {\bibinfo  {journal} {Curr. Opin. Struct. Biol.}\ }\textbf
  {\bibinfo {volume} {21}},\ \bibinfo {pages} {4 -- 11} (\bibinfo {year}
  {2011})}\BibitemShut {NoStop}%
\bibitem [{\citenamefont {Prinz}\ \emph {et~al.}(2011)\citenamefont {Prinz},
  \citenamefont {Wu}, \citenamefont {Sarich}, \citenamefont {Keller},
  \citenamefont {Senne}, \citenamefont {Held}, \citenamefont {Chodera},
  \citenamefont {Sch{\"u}tte},\ and\ \citenamefont {No{\'e}}}]{prinz2011}%
  \BibitemOpen
  \bibfield  {author} {\bibinfo {author} {\bibfnamefont {J.-H.}\ \bibnamefont
  {Prinz}}, \bibinfo {author} {\bibfnamefont {H.}~\bibnamefont {Wu}}, \bibinfo
  {author} {\bibfnamefont {M.}~\bibnamefont {Sarich}}, \bibinfo {author}
  {\bibfnamefont {B.}~\bibnamefont {Keller}}, \bibinfo {author} {\bibfnamefont
  {M.}~\bibnamefont {Senne}}, \bibinfo {author} {\bibfnamefont
  {M.}~\bibnamefont {Held}}, \bibinfo {author} {\bibfnamefont {J.~D.}\
  \bibnamefont {Chodera}}, \bibinfo {author} {\bibfnamefont {C.}~\bibnamefont
  {Sch{\"u}tte}}, \ and\ \bibinfo {author} {\bibfnamefont {F.}~\bibnamefont
  {No{\'e}}},\ }\bibfield  {title} {\enquote {\bibinfo {title} {Markov models
  of molecular kinetics: Generation and validation},}\ }\href@noop {}
  {\bibfield  {journal} {\bibinfo  {journal} {J. Chem. Phys.}\ }\textbf
  {\bibinfo {volume} {134}},\ \bibinfo {pages} {174105} (\bibinfo {year}
  {2011})}\BibitemShut {NoStop}%
\bibitem [{\citenamefont {Dobson}(2003)}]{dobs03}%
  \BibitemOpen
  \bibfield  {author} {\bibinfo {author} {\bibfnamefont {C.~M.}\ \bibnamefont
  {Dobson}},\ }\bibfield  {title} {\enquote {\bibinfo {title} {Protein folding
  and misfolding},}\ }\href {\doibase 10.1038/nature02261} {\bibfield
  {journal} {\bibinfo  {journal} {Nature}\ }\textbf {\bibinfo {volume} {426}},\
  \bibinfo {pages} {884--890} (\bibinfo {year} {2003})}\BibitemShut {NoStop}%
\bibitem [{\citenamefont {No{\'e}}(2015)}]{noe2015}%
  \BibitemOpen
  \bibfield  {author} {\bibinfo {author} {\bibfnamefont {F.}~\bibnamefont
  {No{\'e}}},\ }\bibfield  {title} {\enquote {\bibinfo {title} {Beating the
  millisecond barrier in molecular dynamics simulations},}\ }\href@noop {}
  {\bibfield  {journal} {\bibinfo  {journal} {Biophys. J.}\ }\textbf {\bibinfo
  {volume} {108}},\ \bibinfo {pages} {228--229} (\bibinfo {year}
  {2015})}\BibitemShut {NoStop}%
\bibitem [{\citenamefont {Larson}(2005)}]{lars05}%
  \BibitemOpen
  \bibfield  {author} {\bibinfo {author} {\bibfnamefont {R.~G.}\ \bibnamefont
  {Larson}},\ }\bibfield  {title} {\enquote {\bibinfo {title} {The rheology of
  dilute solutions of flexible polymers: Progress and problems},}\ }\href
  {\doibase 10.1122/1.1835336} {\bibfield  {journal} {\bibinfo  {journal} {J.
  Rheol.}\ }\textbf {\bibinfo {volume} {49}},\ \bibinfo {pages} {1--70}
  (\bibinfo {year} {2005})}\BibitemShut {NoStop}%
\bibitem [{\citenamefont {Smith}\ \emph {et~al.}(1999)\citenamefont {Smith},
  \citenamefont {Babcock},\ and\ \citenamefont {Chu}}]{smit99}%
  \BibitemOpen
  \bibfield  {author} {\bibinfo {author} {\bibfnamefont {D.~E.}\ \bibnamefont
  {Smith}}, \bibinfo {author} {\bibfnamefont {H.~P.}\ \bibnamefont {Babcock}},
  \ and\ \bibinfo {author} {\bibfnamefont {S.}~\bibnamefont {Chu}},\ }\bibfield
   {title} {\enquote {\bibinfo {title} {Single-polymer dynamics in steady shear
  flow},}\ }\href {\doibase 10.1126/science.283.5408.1724} {\bibfield
  {journal} {\bibinfo  {journal} {Science}\ }\textbf {\bibinfo {volume}
  {283}},\ \bibinfo {pages} {1724--1727} (\bibinfo {year} {1999})}\BibitemShut
  {NoStop}%
\bibitem [{\citenamefont {De~Gennes}(1974)}]{genn74}%
  \BibitemOpen
  \bibfield  {author} {\bibinfo {author} {\bibfnamefont {P.~G.}\ \bibnamefont
  {De~Gennes}},\ }\bibfield  {title} {\enquote {\bibinfo {title} {Coil-stretch
  transition of dilute flexible polymers under ultrahigh velocity gradients},}\
  }\href {\doibase 10.1063/1.1681018} {\bibfield  {journal} {\bibinfo
  {journal} {J. Chem. Phys.}\ }\textbf {\bibinfo {volume} {60}},\ \bibinfo
  {pages} {5030--5042} (\bibinfo {year} {1974})}\BibitemShut {NoStop}%
\bibitem [{\citenamefont {Doyle}\ \emph {et~al.}(2000)\citenamefont {Doyle},
  \citenamefont {Ladoux},\ and\ \citenamefont {Viovy}}]{doyl00}%
  \BibitemOpen
  \bibfield  {author} {\bibinfo {author} {\bibfnamefont {P.~S.}\ \bibnamefont
  {Doyle}}, \bibinfo {author} {\bibfnamefont {B.}~\bibnamefont {Ladoux}}, \
  and\ \bibinfo {author} {\bibfnamefont {J.-L.}\ \bibnamefont {Viovy}},\
  }\bibfield  {title} {\enquote {\bibinfo {title} {Dynamics of a tethered
  polymer in shear flow},}\ }\href {\doibase 10.1103/PhysRevLett.84.4769}
  {\bibfield  {journal} {\bibinfo  {journal} {Phys. Rev. Lett.}\ }\textbf
  {\bibinfo {volume} {84}},\ \bibinfo {pages} {4769--4772} (\bibinfo {year}
  {2000})}\BibitemShut {NoStop}%
\bibitem [{\citenamefont {Delgado-Buscalioni}(2006)}]{delgado2006}%
  \BibitemOpen
  \bibfield  {author} {\bibinfo {author} {\bibfnamefont {R.}~\bibnamefont
  {Delgado-Buscalioni}},\ }\bibfield  {title} {\enquote {\bibinfo {title}
  {Cyclic motion of a grafted polymer under shear flow},}\ }\href@noop {}
  {\bibfield  {journal} {\bibinfo  {journal} {Phys. Rev. Lett.}\ }\textbf
  {\bibinfo {volume} {96}},\ \bibinfo {pages} {088303} (\bibinfo {year}
  {2006})}\BibitemShut {NoStop}%
\bibitem [{\citenamefont {Zhang}\ \emph {et~al.}(2009)\citenamefont {Zhang},
  \citenamefont {Donev}, \citenamefont {Weisgraber}, \citenamefont {Alder},
  \citenamefont {Graham},\ and\ \citenamefont {de~Pablo}}]{zhan09}%
  \BibitemOpen
  \bibfield  {author} {\bibinfo {author} {\bibfnamefont {Y.}~\bibnamefont
  {Zhang}}, \bibinfo {author} {\bibfnamefont {A.}~\bibnamefont {Donev}},
  \bibinfo {author} {\bibfnamefont {T.}~\bibnamefont {Weisgraber}}, \bibinfo
  {author} {\bibfnamefont {B.~J.}\ \bibnamefont {Alder}}, \bibinfo {author}
  {\bibfnamefont {M.~D.}\ \bibnamefont {Graham}}, \ and\ \bibinfo {author}
  {\bibfnamefont {J.~J.}\ \bibnamefont {de~Pablo}},\ }\bibfield  {title}
  {\enquote {\bibinfo {title} {Tethered dna dynamics in shear flow},}\ }\href
  {\doibase 10.1063/1.3149860} {\bibfield  {journal} {\bibinfo  {journal} {The
  Journal of Chemical Physics}\ }\textbf {\bibinfo {volume} {130}},\ \bibinfo
  {pages} {234902} (\bibinfo {year} {2009})}\BibitemShut {NoStop}%
\bibitem [{\citenamefont {Wang}\ and\ \citenamefont
  {Sch{\"u}tte}(2015)}]{wang2015}%
  \BibitemOpen
  \bibfield  {author} {\bibinfo {author} {\bibfnamefont {H.}~\bibnamefont
  {Wang}}\ and\ \bibinfo {author} {\bibfnamefont {C.}~\bibnamefont
  {Sch{\"u}tte}},\ }\bibfield  {title} {\enquote {\bibinfo {title} {Building
  markov state models for periodically driven non-equilibrium systems},}\
  }\href@noop {} {\bibfield  {journal} {\bibinfo  {journal} {J. Chem. Theory
  Comput.}\ }\textbf {\bibinfo {volume} {11}},\ \bibinfo {pages} {1819--1831}
  (\bibinfo {year} {2015})}\BibitemShut {NoStop}%
\bibitem [{\citenamefont {Koltai}\ \emph {et~al.}(2016)\citenamefont {Koltai},
  \citenamefont {Ciccotti},\ and\ \citenamefont {Sch{\"u}tte}}]{koltai2016}%
  \BibitemOpen
  \bibfield  {author} {\bibinfo {author} {\bibfnamefont {P.}~\bibnamefont
  {Koltai}}, \bibinfo {author} {\bibfnamefont {G.}~\bibnamefont {Ciccotti}}, \
  and\ \bibinfo {author} {\bibfnamefont {C.}~\bibnamefont {Sch{\"u}tte}},\
  }\href@noop {} {\emph {\bibinfo {title} {On Markov state models for
  non-equilibrium molecular dynamics}}},\ \bibinfo {type} {Tech. Rep.}\
  \bibinfo {number} {16-11}\ (\bibinfo  {institution} {ZIB},\ \bibinfo
  {address} {Takustr.7, 14195 Berlin},\ \bibinfo {year} {2016})\BibitemShut
  {NoStop}%
\bibitem [{\citenamefont {Pellegrini}\ \emph {et~al.}(2016)\citenamefont
  {Pellegrini}, \citenamefont {Landes}, \citenamefont {Laio}, \citenamefont
  {Prestipino},\ and\ \citenamefont {Tosatti}}]{pelle16}%
  \BibitemOpen
  \bibfield  {author} {\bibinfo {author} {\bibfnamefont {F.}~\bibnamefont
  {Pellegrini}}, \bibinfo {author} {\bibfnamefont {F.~P.}\ \bibnamefont
  {Landes}}, \bibinfo {author} {\bibfnamefont {A.}~\bibnamefont {Laio}},
  \bibinfo {author} {\bibfnamefont {S.}~\bibnamefont {Prestipino}}, \ and\
  \bibinfo {author} {\bibfnamefont {E.}~\bibnamefont {Tosatti}},\ }\bibfield
  {title} {\enquote {\bibinfo {title} {Markov state modeling of sliding
  friction},}\ }\href@noop {} {\bibfield  {journal} {\bibinfo  {journal}
  {arXiv:1605.03849}\ } (\bibinfo {year} {2016})}\BibitemShut {NoStop}%
\bibitem [{\citenamefont {Puglisi}\ \emph {et~al.}(2010)\citenamefont
  {Puglisi}, \citenamefont {Pigolotti}, \citenamefont {Rondoni},\ and\
  \citenamefont {Vulpiani}}]{pugl10}%
  \BibitemOpen
  \bibfield  {author} {\bibinfo {author} {\bibfnamefont {A.}~\bibnamefont
  {Puglisi}}, \bibinfo {author} {\bibfnamefont {S.}~\bibnamefont {Pigolotti}},
  \bibinfo {author} {\bibfnamefont {L.}~\bibnamefont {Rondoni}}, \ and\
  \bibinfo {author} {\bibfnamefont {A.}~\bibnamefont {Vulpiani}},\ }\bibfield
  {title} {\enquote {\bibinfo {title} {Entropy production and coarse graining
  in markov processes},}\ }\href {\doibase 10.1088/1742-5468/2010/05/P05015}
  {\bibfield  {journal} {\bibinfo  {journal} {J. Stat. Mech.}\ ,\ \bibinfo
  {pages} {P05015}} (\bibinfo {year} {2010})}\BibitemShut {NoStop}%
\bibitem [{\citenamefont {Knoch}\ and\ \citenamefont
  {Speck}(2015)}]{knoch2015}%
  \BibitemOpen
  \bibfield  {author} {\bibinfo {author} {\bibfnamefont {F.}~\bibnamefont
  {Knoch}}\ and\ \bibinfo {author} {\bibfnamefont {T.}~\bibnamefont {Speck}},\
  }\bibfield  {title} {\enquote {\bibinfo {title} {Cycle representatives for
  the coarse-graining of systems driven into a non-equilibrium steady state},}\
  }\href@noop {} {\bibfield  {journal} {\bibinfo  {journal} {New J. Phys.}\
  }\textbf {\bibinfo {volume} {17}},\ \bibinfo {pages} {115004} (\bibinfo
  {year} {2015})}\BibitemShut {NoStop}%
\bibitem [{\citenamefont {Seifert}(2012)}]{seif12}%
  \BibitemOpen
  \bibfield  {author} {\bibinfo {author} {\bibfnamefont {U.}~\bibnamefont
  {Seifert}},\ }\bibfield  {title} {\enquote {\bibinfo {title} {Stochastic
  thermodynamics, fluctuation theorems, and molecular machines},}\ }\href
  {\doibase doi:10.1088/0034-4885/75/12/126001} {\bibfield  {journal} {\bibinfo
   {journal} {Rep. Prog. Phys.}\ }\textbf {\bibinfo {volume} {75}},\ \bibinfo
  {pages} {126001} (\bibinfo {year} {2012})}\BibitemShut {NoStop}%
\bibitem [{\citenamefont {Alexander-Katz}\ \emph {et~al.}(2006)\citenamefont
  {Alexander-Katz}, \citenamefont {Schneider}, \citenamefont {Schneider},
  \citenamefont {Wixforth},\ and\ \citenamefont {Netz}}]{alex2006}%
  \BibitemOpen
  \bibfield  {author} {\bibinfo {author} {\bibfnamefont {A.}~\bibnamefont
  {Alexander-Katz}}, \bibinfo {author} {\bibfnamefont {M.~F.}\ \bibnamefont
  {Schneider}}, \bibinfo {author} {\bibfnamefont {S.~W.}\ \bibnamefont
  {Schneider}}, \bibinfo {author} {\bibfnamefont {A.}~\bibnamefont {Wixforth}},
  \ and\ \bibinfo {author} {\bibfnamefont {R.~R.}\ \bibnamefont {Netz}},\
  }\bibfield  {title} {\enquote {\bibinfo {title} {Shear-flow-induced unfolding
  of polymeric globules},}\ }\href@noop {} {\bibfield  {journal} {\bibinfo
  {journal} {Phys. Rev. Lett.}\ }\textbf {\bibinfo {volume} {97}},\ \bibinfo
  {pages} {138101} (\bibinfo {year} {2006})}\BibitemShut {NoStop}%
\bibitem [{\citenamefont {Winkler}(2006)}]{wink06}%
  \BibitemOpen
  \bibfield  {author} {\bibinfo {author} {\bibfnamefont {R.~G.}\ \bibnamefont
  {Winkler}},\ }\bibfield  {title} {\enquote {\bibinfo {title} {Semiflexible
  polymers in shear flow},}\ }\href {\doibase 10.1103/PhysRevLett.97.128301}
  {\bibfield  {journal} {\bibinfo  {journal} {Phys. Rev. Lett.}\ }\textbf
  {\bibinfo {volume} {97}},\ \bibinfo {pages} {128301} (\bibinfo {year}
  {2006})}\BibitemShut {NoStop}%
\bibitem [{\citenamefont {Lemak}\ \emph {et~al.}(1998)\citenamefont {Lemak},
  \citenamefont {Balabaev}, \citenamefont {Karnet},\ and\ \citenamefont
  {Yanovsky}}]{lemak1998}%
  \BibitemOpen
  \bibfield  {author} {\bibinfo {author} {\bibfnamefont {A.~S.}\ \bibnamefont
  {Lemak}}, \bibinfo {author} {\bibfnamefont {N.~K.}\ \bibnamefont {Balabaev}},
  \bibinfo {author} {\bibfnamefont {Y.~N.}\ \bibnamefont {Karnet}}, \ and\
  \bibinfo {author} {\bibfnamefont {Y.~G.}\ \bibnamefont {Yanovsky}},\
  }\bibfield  {title} {\enquote {\bibinfo {title} {The effect of a solid wall
  on polymer chain behavior under shear flow},}\ }\href@noop {} {\bibfield
  {journal} {\bibinfo  {journal} {J. Chem. Phys.}\ }\textbf {\bibinfo {volume}
  {108}},\ \bibinfo {pages} {797--806} (\bibinfo {year} {1998})}\BibitemShut
  {NoStop}%
\bibitem [{\citenamefont {Altaner}\ \emph {et~al.}(2012)\citenamefont
  {Altaner}, \citenamefont {Grosskinsky}, \citenamefont {Herminghaus},
  \citenamefont {Katth\"an}, \citenamefont {Timme},\ and\ \citenamefont
  {Vollmer}}]{alta12a}%
  \BibitemOpen
  \bibfield  {author} {\bibinfo {author} {\bibfnamefont {B.}~\bibnamefont
  {Altaner}}, \bibinfo {author} {\bibfnamefont {S.}~\bibnamefont
  {Grosskinsky}}, \bibinfo {author} {\bibfnamefont {S.}~\bibnamefont
  {Herminghaus}}, \bibinfo {author} {\bibfnamefont {L.}~\bibnamefont
  {Katth\"an}}, \bibinfo {author} {\bibfnamefont {M.}~\bibnamefont {Timme}}, \
  and\ \bibinfo {author} {\bibfnamefont {J.}~\bibnamefont {Vollmer}},\
  }\bibfield  {title} {\enquote {\bibinfo {title} {Network representations of
  nonequilibrium steady states: Cycle decompositions, symmetries, and dominant
  paths},}\ }\href {\doibase 10.1103/PhysRevE.85.041133} {\bibfield  {journal}
  {\bibinfo  {journal} {Phys. Rev. E}\ }\textbf {\bibinfo {volume} {85}},\
  \bibinfo {pages} {041133} (\bibinfo {year} {2012})}\BibitemShut {NoStop}%
\bibitem [{\citenamefont {Kalpazidou}(2007)}]{kalpazidou2007}%
  \BibitemOpen
  \bibfield  {author} {\bibinfo {author} {\bibfnamefont {S.~L.}\ \bibnamefont
  {Kalpazidou}},\ }\href@noop {} {\emph {\bibinfo {title} {Cycle
  representations of Markov processes}}},\ Vol.~\bibinfo {volume} {28}\
  (\bibinfo  {publisher} {Springer Science \& Business Media},\ \bibinfo {year}
  {2007})\BibitemShut {NoStop}%
\bibitem [{\citenamefont {Schnakenberg}(1976)}]{schnakenberg1976}%
  \BibitemOpen
  \bibfield  {author} {\bibinfo {author} {\bibfnamefont {J.}~\bibnamefont
  {Schnakenberg}},\ }\bibfield  {title} {\enquote {\bibinfo {title} {Network
  theory of microscopic and macroscopic behavior of master equation systems},}\
  }\href@noop {} {\bibfield  {journal} {\bibinfo  {journal} {Rev. Mod. Phys.}\
  }\textbf {\bibinfo {volume} {48}},\ \bibinfo {pages} {571} (\bibinfo {year}
  {1976})}\BibitemShut {NoStop}%
\bibitem [{\citenamefont {Conrad}\ \emph {et~al.}(2015)\citenamefont {Conrad},
  \citenamefont {Banisch},\ and\ \citenamefont
  {Sch{\"u}tte}}]{conrad2014modularity}%
  \BibitemOpen
  \bibfield  {author} {\bibinfo {author} {\bibfnamefont {N.~D.}\ \bibnamefont
  {Conrad}}, \bibinfo {author} {\bibfnamefont {R.}~\bibnamefont {Banisch}}, \
  and\ \bibinfo {author} {\bibfnamefont {C.}~\bibnamefont {Sch{\"u}tte}},\
  }\bibfield  {title} {\enquote {\bibinfo {title} {Modularity of directed
  networks: Cycle decomposition approach},}\ }\href {\doibase
  10.3934/jcd.2015.2.1} {\bibfield  {journal} {\bibinfo  {journal} {J. Comput.
  Dyn.}\ }\textbf {\bibinfo {volume} {2}},\ \bibinfo {pages} {1--24} (\bibinfo
  {year} {2015})}\BibitemShut {NoStop}%
\bibitem [{Note1()}]{Note1}%
  \BibitemOpen
  \bibinfo {note} {The reason why the connectivity fails is presumably because
  the underlying potential energy landscape does not offer well separated
  basins.}\BibitemShut {Stop}%
\bibitem [{\citenamefont {Jolliffe}(2002)}]{jolli2002}%
  \BibitemOpen
  \bibfield  {author} {\bibinfo {author} {\bibfnamefont {I.}~\bibnamefont
  {Jolliffe}},\ }\href@noop {} {\emph {\bibinfo {title} {Principal component
  analysis}}}\ (\bibinfo  {publisher} {Wiley Online Library},\ \bibinfo {year}
  {2002})\BibitemShut {NoStop}%
\bibitem [{cme()}]{cmeans}%
  \BibitemOpen
  \href@noop {} {}\bibinfo {note}
  {\text{http://pythonhosted.org/scikit-fuzzy}}\BibitemShut {NoStop}%
\bibitem [{\citenamefont {Altaner}\ and\ \citenamefont
  {Vollmer}(2012)}]{alta2012}%
  \BibitemOpen
  \bibfield  {author} {\bibinfo {author} {\bibfnamefont {B.}~\bibnamefont
  {Altaner}}\ and\ \bibinfo {author} {\bibfnamefont {J.}~\bibnamefont
  {Vollmer}},\ }\bibfield  {title} {\enquote {\bibinfo {title}
  {Fluctuation-preserving coarse graining for biochemical systems},}\
  }\href@noop {} {\bibfield  {journal} {\bibinfo  {journal} {Phys. Rev. Lett.}\
  }\textbf {\bibinfo {volume} {108}},\ \bibinfo {pages} {228101} (\bibinfo
  {year} {2012})}\BibitemShut {NoStop}%
\bibitem [{\citenamefont {England}(2013)}]{engl13}%
  \BibitemOpen
  \bibfield  {author} {\bibinfo {author} {\bibfnamefont {J.~L.}\ \bibnamefont
  {England}},\ }\bibfield  {title} {\enquote {\bibinfo {title} {Statistical
  physics of self-replication},}\ }\href {\doibase 10.1063/1.4818538}
  {\bibfield  {journal} {\bibinfo  {journal} {J. Chem. Phys.}\ }\textbf
  {\bibinfo {volume} {139}},\ \bibinfo {pages} {121923} (\bibinfo {year}
  {2013})}\BibitemShut {NoStop}%
\bibitem [{\citenamefont {Weber}\ \emph {et~al.}(2015)\citenamefont {Weber},
  \citenamefont {Shukla},\ and\ \citenamefont {Pande}}]{weber2015}%
  \BibitemOpen
  \bibfield  {author} {\bibinfo {author} {\bibfnamefont {J.~K.}\ \bibnamefont
  {Weber}}, \bibinfo {author} {\bibfnamefont {D.}~\bibnamefont {Shukla}}, \
  and\ \bibinfo {author} {\bibfnamefont {V.~S.}\ \bibnamefont {Pande}},\
  }\bibfield  {title} {\enquote {\bibinfo {title} {Heat dissipation guides
  activation in signaling proteins},}\ }\href@noop {} {\bibfield  {journal}
  {\bibinfo  {journal} {Proc. Natl. Acad. Sci. USA}\ }\textbf {\bibinfo
  {volume} {112}},\ \bibinfo {pages} {10377--10382} (\bibinfo {year}
  {2015})}\BibitemShut {NoStop}%
\bibitem [{\citenamefont {P{\'e}rez-Hern{\'a}ndez}\ \emph
  {et~al.}(2013)\citenamefont {P{\'e}rez-Hern{\'a}ndez}, \citenamefont {Paul},
  \citenamefont {Giorgino}, \citenamefont {De~Fabritiis},\ and\ \citenamefont
  {No{\'e}}}]{perez2013}%
  \BibitemOpen
  \bibfield  {author} {\bibinfo {author} {\bibfnamefont {G.}~\bibnamefont
  {P{\'e}rez-Hern{\'a}ndez}}, \bibinfo {author} {\bibfnamefont
  {F.}~\bibnamefont {Paul}}, \bibinfo {author} {\bibfnamefont {T.}~\bibnamefont
  {Giorgino}}, \bibinfo {author} {\bibfnamefont {G.}~\bibnamefont
  {De~Fabritiis}}, \ and\ \bibinfo {author} {\bibfnamefont {F.}~\bibnamefont
  {No{\'e}}},\ }\bibfield  {title} {\enquote {\bibinfo {title} {Identification
  of slow molecular order parameters for markov model construction},}\
  }\href@noop {} {\bibfield  {journal} {\bibinfo  {journal} {J. Chem. Phys.}\
  }\textbf {\bibinfo {volume} {139}},\ \bibinfo {pages} {015102} (\bibinfo
  {year} {2013})}\BibitemShut {NoStop}%
\bibitem [{\citenamefont {Barato}\ and\ \citenamefont
  {Seifert}(2015)}]{bara15}%
  \BibitemOpen
  \bibfield  {author} {\bibinfo {author} {\bibfnamefont {A.~C.}\ \bibnamefont
  {Barato}}\ and\ \bibinfo {author} {\bibfnamefont {U.}~\bibnamefont
  {Seifert}},\ }\bibfield  {title} {\enquote {\bibinfo {title} {Thermodynamic
  uncertainty relation for biomolecular processes},}\ }\href {\doibase
  10.1103/PhysRevLett.114.158101} {\bibfield  {journal} {\bibinfo  {journal}
  {Phys. Rev. Lett.}\ }\textbf {\bibinfo {volume} {114}},\ \bibinfo {pages}
  {158101} (\bibinfo {year} {2015})}\BibitemShut {NoStop}%
\bibitem [{\citenamefont {Gingrich}\ \emph {et~al.}(2016)\citenamefont
  {Gingrich}, \citenamefont {Horowitz}, \citenamefont {Perunov},\ and\
  \citenamefont {England}}]{ging16}%
  \BibitemOpen
  \bibfield  {author} {\bibinfo {author} {\bibfnamefont {T.~R.}\ \bibnamefont
  {Gingrich}}, \bibinfo {author} {\bibfnamefont {J.~M.}\ \bibnamefont
  {Horowitz}}, \bibinfo {author} {\bibfnamefont {N.}~\bibnamefont {Perunov}}, \
  and\ \bibinfo {author} {\bibfnamefont {J.~L.}\ \bibnamefont {England}},\
  }\bibfield  {title} {\enquote {\bibinfo {title} {Dissipation bounds all
  steady-state current fluctuations},}\ }\href {\doibase
  10.1103/PhysRevLett.116.120601} {\bibfield  {journal} {\bibinfo  {journal}
  {Phys. Rev. Lett.}\ }\textbf {\bibinfo {volume} {116}},\ \bibinfo {pages}
  {120601} (\bibinfo {year} {2016})}\BibitemShut {NoStop}%
\bibitem [{\citenamefont {Sugawa}\ \emph {et~al.}(2016)\citenamefont {Sugawa},
  \citenamefont {Okazaki}, \citenamefont {Kobayashi}, \citenamefont {Matsui},
  \citenamefont {Hummer}, \citenamefont {Masaike},\ and\ \citenamefont
  {Nishizaka}}]{sugawa2016}%
  \BibitemOpen
  \bibfield  {author} {\bibinfo {author} {\bibfnamefont {M.}~\bibnamefont
  {Sugawa}}, \bibinfo {author} {\bibfnamefont {K.}~\bibnamefont {Okazaki}},
  \bibinfo {author} {\bibfnamefont {M.}~\bibnamefont {Kobayashi}}, \bibinfo
  {author} {\bibfnamefont {T.}~\bibnamefont {Matsui}}, \bibinfo {author}
  {\bibfnamefont {G.}~\bibnamefont {Hummer}}, \bibinfo {author} {\bibfnamefont
  {T.}~\bibnamefont {Masaike}}, \ and\ \bibinfo {author} {\bibfnamefont
  {T.}~\bibnamefont {Nishizaka}},\ }\bibfield  {title} {\enquote {\bibinfo
  {title} {F1-atpase conformational cycle from simultaneous single-molecule
  fret and rotation measurements},}\ }\href {\doibase 10.1073/pnas.1524720113}
  {\bibfield  {journal} {\bibinfo  {journal} {Proc. Natl. Acad. Sci. USA}\
  }\textbf {\bibinfo {volume} {113}},\ \bibinfo {pages} {E2916--E2924}
  (\bibinfo {year} {2016})}\BibitemShut {NoStop}%
\bibitem [{\citenamefont {Swope}\ \emph {et~al.}(2004)\citenamefont {Swope},
  \citenamefont {Pitera},\ and\ \citenamefont {Suits}}]{swo2004}%
  \BibitemOpen
  \bibfield  {author} {\bibinfo {author} {\bibfnamefont {W.~C.}\ \bibnamefont
  {Swope}}, \bibinfo {author} {\bibfnamefont {J.~W.}\ \bibnamefont {Pitera}}, \
  and\ \bibinfo {author} {\bibfnamefont {F.}~\bibnamefont {Suits}},\ }\bibfield
   {title} {\enquote {\bibinfo {title} {Describing protein folding kinetics by
  molecular dynamics simulations. 1. theory},}\ }\href@noop {} {\bibfield
  {journal} {\bibinfo  {journal} {J. Phys. Chem. B}\ }\textbf {\bibinfo
  {volume} {108}},\ \bibinfo {pages} {6571--6581} (\bibinfo {year}
  {2004})}\BibitemShut {NoStop}%
\bibitem [{\citenamefont {Newman}(2010)}]{newman2010}%
  \BibitemOpen
  \bibfield  {author} {\bibinfo {author} {\bibfnamefont {M.}~\bibnamefont
  {Newman}},\ }\href@noop {} {\emph {\bibinfo {title} {Networks: An
  Introduction}}}\ (\bibinfo  {publisher} {Oxford university press},\ \bibinfo
  {year} {2010})\BibitemShut {NoStop}%
\end{thebibliography}
\end{document}